\documentclass[pra,superscriptaddress,twocolumn,showpacs,10pt]{revtex4}

\usepackage{amssymb}
\usepackage{amsmath}
\usepackage{graphicx}

\newtheorem{theorem}{Theorem}
\newtheorem{corollary}{Corollary}
\newtheorem{example}{Example}

\begin{document}

\title{Deterministic distributed dense coding with stabilizer states}
\author{Guoming Wang}
\email{wgm00@mails.tsinghua.edu.cn}
\author{Mingsheng Ying}
\email{yingmsh@tsinghua.edu.cn}

\affiliation{State Key Laboratory of Intelligent Technology and
Systems, Department of Computer Science and Technology, Tsinghua
University, Beijing, China, 100084}

\date{\today}

\begin{abstract}
We consider the possibility of using stabilizer states to perform
deterministic dense coding among multiple senders and a single
receiver. In the model we studied, the utilized stabilizer state
is partitioned into several subsystems and then
each subsystem is held by a distinct party.
We present a sufficient condition for a
stabilizer state to be useful for deterministic distributed dense
coding with respect to a given partition plan. The corresponding
protocol is also constructed. Furthermore, we propose a method to
partially solve a more general problem of finding the set of
achievable alphabet sizes for an arbitrary stabilizer state with
respect to an arbitrary partition plan. Finally, our work provides
a new perspective from the stabilizer formalism to view the
standard dense coding protocol and also unifies several previous
results in a single framework.
\end{abstract}


\maketitle

\section{Introduction}

Since its proposal by Bennett and Wiesner in 1992 \cite{BW92},
dense coding has become one of the most important constituents in
quantum information science. This communication protocol enables
enhancement of the classical capacity of a noiseless quantum
channel by using previously shared entanglement between the sender
and the receiver. Up to now, researchers are still trying to
thoroughly understand the power of a general bipartite entangled
state in this task
\cite{B01,M02,MO05,PP05,JF06,WC06,FD06,FZ06,BG07}. Typically there
are two classes of dense coding schemes considered. One is called
deterministic dense coding, which requires the protocol to succeed
all the time; while the other one, performing unambiguous
discrimination \cite{IV87,DI88,PE88} on the final state, allows
the protocol to succeed in a probabilistic manner.

Recently several authors have begun to consider the possibility of
using a multipartite entangled state to perform dense coding among
multiple parties \cite{LL02,LA02,GW02,BD04,BD05,R04,AP06,
YC06,HP07,PA07,MP07}. In the multipartite case, many senders may
\textit{simultaneously} transmit classical information to a single
receiver with the aid of \textit{a priori} multipartite
entanglement. Since each sender can only encode on his own
subsystem, this scheme is called `distributed dense
coding'\cite{BD04,BD05}. Specifically, our model of deterministic
distributed dense coding is as follows. Suppose $\rho$ is an
$n$-qudit state. Divide its $n$ qudits into $m$ groups
$T_1,T_2,\dots,T_m$ for some $2 \le m \le n$ and then distribute
the subsystem $T_i$ to the $i$-th party $A_i$, for
$i=1,2,\dots,m$. Now assume that $A_i$ performs one out of $b_i$
different quantum operations on the subsystem $T_i$, for
$i=1,2,\dots,m-1$. Then $A_1,A_2,\dots,A_{m-1}$ send all their
subsystems to $A_{m}$. If $A_{m}$ can perfectly distinguish among
all possible states, then this procedure actually accomplishes
transmission of $log_2b_i$ bits of classical information from
$A_i$ to $A_m$, for $i=1,2,\dots,m-1$. In this case, we say that
$(b_1,b_2,\dots,b_{m-1})$ is an achievable alphabet size for
$\rho$ with respect to the grouping plan $T_1,T_2,\dots,T_m$. Then
for a given state $\rho$, any grouping strategy will define a
region of achievable alphabet sizes. The most general question
would be to determine such a region for all possible partition
plans. For a more practical concern, we want to know whether the
utilization of $\rho$ really improves the classical capacity of
the senders. So only when there exists an achievable alphabet size
$(b_1,b_2,\dots,b_{m-1})$ with $b_i>d^{|T_i|}$ for at least one $1
\le i \le m-1$ and $b_j \ge d^{|T_j|}$ for other $j \neq i$ (where
$|T_i|$ denotes the number of qudits in $T_i$), we say that $\rho$
is useful for deterministic distributed dense coding with respect
to $T_1,T_2,\dots,T_m$.

The purpose of this paper is to investigate the usefulness of
stabilizer states for deterministic distributed dense coding.
Stabilizer states have played an important role in quantum
information theory, especially in the field of quantum error
correction \cite{S95,S96} and cluster state quantum computation
\cite{RB01}. They can be described in an elegant and compact form
named the stabilizer formalism \cite{G96,G97}, which has also lead
to novel perspectives to many phenomena in quantum information
science and quantum mechanics \cite{TG05,ND07,WY07}. We present a
sufficient condition for a stabilizer state to be useful for
deterministic distributed dense coding with respect to a given
partition plan. The corresponding protocol is also constructed.
Furthermore, we propose a method to partially solve the general
problem of finding the region of achievable alphabet sizes for an
arbitrary stabilizer state with respect to an arbitrary partition
plan. Finally, our work provides a new perspective from the
stabilizer formalism to view the standard dense coding protocol
and also unifies several previous results in a single framework.

This paper is organized as follows. In Sec. II we briefly recall
some fundamental facts about the stabilizer formalism. In Sec.
III, we study the power of stabilizer states in deterministic
distributed dense coding and also construct the corresponding
protocol. In Sec. IV we analyze several concrete examples by using
our theorems. Finally, Sec. V summarizes our results.

\section{Preliminary}

In this section, we review some fundamental facts about stabilizer
state and its corresponding stabilizer formalism. Although in most
literatures the notion of stabilizer state was put forward in the
context of multiqubit systems, it can actually be generalized
without essential difficulty to arbitrary higher-dimensional
systems as well. Similar topics have also been explored in Refs.
\cite{G98,NB02,Y02,HD05}. So here we directly start with the
general higher-dimensional case.

Consider a $d$-dimensional Hilbert space. Define
\begin{equation}
\begin{array}{l}
X^{(d)}=\sum\limits_{j=0}^{d-1}{|j \oplus 1\rangle\langle j|},\\
Z^{(d)}=\sum\limits_{j=0}^{d-1}{\omega^{j}|j\rangle\langle j|},
\end{array}
\label{equ:pauli}
\end{equation}
where $\omega=e^{i\frac{2\pi}{d}}$ is the $d$-th root of unity
over the complex field and the `$\oplus$' sign denotes addition
modulo $d$. In what follows, without causing ambiguity, we will
omit the superscript `$(d)$' in $X^{(d)}$ and $Z^{(d)}$. The
matrices $\{\sigma_{a,b}=X^aZ^b:a,b= 0,1,\dots,d-1\}$ are
considered as the generalized Pauli matrices over $d$-dimensional
space. The commutation relations among them are given by
\begin{equation}
\sigma_{a,b}\sigma_{j,k}=\omega^{bj-ak}\sigma_{j,k}\sigma_{a,b}.
\label{equ:commuterelation}
\end{equation}
It can be checked that if $d$ is even and $ab$ is odd, the
eigenvalues of $\sigma_{a,b}$ are $\omega^{1/2}, \omega^{c+1/2},
\omega^{2c+1/2}, \dots, \omega^{d-c+1/2}$ for some factor $c$ of
$d$; otherwise, the eigenvalues of $\sigma_{a,b}$ are $1,
\omega^c, \omega^{2c}, \dots, \omega^{d-c}$ for some factor $c$ of
$d$.

The generalized Pauli group on $n$ qudits $G^{(d)}_n$ is defined
to consist all $n$-fold tensor products of generalized Pauli
matrices over $d$-dimensional space, allowing overall phase factor
$\gamma^a$, where $\gamma=\sqrt{\omega}$ and $0 \le a \le 2d-1$,
i.e.
\begin{equation}
\begin{array}{ll}
G^{(d)}_n&=\{\gamma^{a}\sigma_{i_1,j_1}\otimes\sigma_{i_2,j_2}\otimes\dots\otimes\sigma_{i_n,j_n}:
0 \le a \le 2d-1,\\
&0 \le i_1,j_1,i_2,j_2,\dots,i_n,j_n \le d-1\}.
\label{equ:pauligroup}
\end{array} \end{equation}
Actually, when $d$ is odd, the introduction of $\gamma$ is
unnecessary and it can be replaced by $\omega$. For a detailed
discussion about this, one can see Ref. \cite{HD05}.

Define the map $\chi: G^{(d)}_{n} \rightarrow \mathbb{Z}_d^{2n}$
as follows: for $g=\gamma^c \sigma_{a_1,b_1}\otimes
\sigma_{a_2,b_2} \otimes\dots \otimes \sigma_{a_n,b_n}$,
$\chi(g)=(a_1,a_2,\dots,a_n,b_1,b_2,\dots,b_n)$. From now on, all
additions and multiplications of $\chi(g)$ will be taken over
$\mathbb{Z}_d$. By Eq.(\ref{equ:commuterelation}), for any
$g=\gamma^c \sigma_{a_1,b_1}\otimes \sigma_{a_2,b_2} \otimes \dots
\otimes \sigma_{a_n,b_n}$, $h=\gamma^{c'}
\sigma_{a'_1,b'_1}\otimes \sigma_{a'_2,b'_2} \otimes \dots \otimes
\sigma_{a'_n,b'_n} \in G^{(d)}_n$, their commutation relation is
\begin{equation}\begin{array}{l}
gh=\omega^{\sum_{i=1}^{n}(b_ia'_i-a_ib'_i)}hg
=\omega^{\chi(g)\Lambda_n \chi(h)^{T}}hg, \label{equ:ghhg}
\end{array}\end{equation}
where $\Lambda_n$ is a $2n \times 2n$ matrix given by
\begin{equation}
\Lambda_n=\begin{pmatrix}
0 & -I_n \\
I_n & 0 \\
\end{pmatrix}
\end{equation}
and $I_n$ is the $n \times n$ identity matrix. So we have
\begin{equation}
ghg^{\dagger}=\omega^{\chi(g)\Lambda_n \chi(h)^{T}}h.
\label{equ:ghg}
\end{equation}
In particular, $g$ and $h$ commute if and only if
\begin{equation}\begin{array}{l}
\chi(g)\Lambda_n \chi(h)^{T}=0.
\end{array}\end{equation}

For a set of commuting operators $g_1,g_2,\dots,g_k \in
G^{(d)}_n$, we say that they are independent if $\forall
i=1,2,\dots,k$,
\begin{equation}
\langle g_1,g_2,\dots,g_k\rangle \neq \langle
g_1,g_2,\dots,g_{i-1},g_{i+1},\dots, g_k\rangle.
\end{equation}
Define $G'^{(d)}_n$ to be the subset of $G^{(d)}_n$ composed of
all the operators whose eigenvalues are of the form $1, \omega^c,
\omega^{2c}, \dots, \omega^{d-c}$ for some factor $c$ of $d$. Now
suppose $g_1,g_2,\dots,g_n$ are independent commuting operators in
$G'^{(d)}_n$. Let
\begin{equation}
S=\langle g_1,g_2,\dots,g_n \rangle
\end{equation}
be the Abelian subgroup generated by them. If there exists a
unique state $|\psi_S\rangle$ (up to an overall phase) such that
\begin{equation}\begin{array}{ll}
g_i|\psi_S\rangle=|\psi_S\rangle, & \forall i=1,2,\dots,n,
\end{array}\end{equation}
we say that $S$ is a \textit{complete} stabilizer and $|\psi_S\rangle$
is stabilized by $S$. In this case, with the fact $\sum_{j=0}^{d-1}{\omega^{j\lambda}}=0$, $\forall
\lambda=1,2,\dots,d-1$, one can verify
\begin{equation}
\rho_S \equiv
|\psi_S\rangle\langle\psi_S|=\frac{1}{d^n}\prod\limits_{i=1}^{n}(\sum\limits_{j=0}^{d-1}{g^j_i}).
\label{equ:ps}
\end{equation}

Suppose $S=\langle g_1,g_2,\dots,g_k\rangle$, where
$g_1,g_2,\dots,g_k$ are independent commuting operations in
$G'^{(d)}_n$. There is an extremely useful way of presenting the
generators $g_1,g_2,\dots,g_k$ using the check matrix $M$. This
matrix is of size $k \times 2n$ and its $i$-th row is simply the
representation row of the $i$-th generator $\chi(g_i)$, $\forall
i=1,2,\dots,k$. Since $g_1,g_2,\dots,g_k$ mutually commute, the
check matrix $M$ satisfies
\begin{equation}\begin{array}{l}
M \Lambda_n M^{\dagger}=0.
\end{array}\end{equation}
For example, consider a four-qutrit system, i.e. $d=3$, $n=4$.
\begin{equation} \begin{array}{l}
g_1=\sigma_{1,0} \otimes \sigma_{0,1} \otimes \sigma_{1,1} \otimes \sigma_{0,1}, \\
g_2=\sigma_{0,2} \otimes \sigma_{2,0} \otimes \sigma_{0,1} \otimes \sigma_{1,1},\\
g_3=\sigma_{1,1} \otimes \sigma_{0,1} \otimes \sigma_{0,2} \otimes
\sigma_{0,0}
\end{array} \end{equation}
are three independent commuting operators from $G'^{(3)}_4$. Then
the corresponding check matrix is
\begin{equation}
M=\begin{pmatrix}
1 & 0 & 1 & 0  & 0 & 1 & 1 & 1\\
0 & 2 & 0 & 1  & 2 & 0 & 1 & 1\\
1 & 0 & 0 & 0  & 1 & 1 & 2 & 0\\
\end{pmatrix}.
\end{equation}

\section{Deterministic dense coding with stabilizer states}

In this section we investigate the usefulness of stabilizer states
for deterministic distributed dense coding.

At first, we need to introduce two groups of definitions and
notations. The first group is about sets of integers. We use
$[1,n]$ to denote the set of integers $\{1,2,\dots,n\}$. If
$T_1,T_2,\dots,T_k$ are disjoint proper subsets of $[1,n]$ and
they satisfy $\cup_{i=1}^{k} T_i=[1,n]$, then we say
$(T_1,T_2,\dots,T_k)$ a partition of $[1,n]$. We also use $|T|$ to
denote the number of elements in a set $T$. The second group is
about vectors in $\mathbb{Z}^n_d$. Note that all additions and
multiplications of vectors in $\mathbb{Z}^n_d$ are taken over
$\mathbb{Z}_d$. For any
$\overrightarrow{\alpha}_1,\overrightarrow{\alpha}_2,\dots,\overrightarrow{\alpha}_k
\in \mathbb{Z}^n_d$, their linear span is defined as
\begin{equation} \begin{array}{l}
span\{\overrightarrow{\alpha}_1,\overrightarrow{\alpha}_2,\dots,\overrightarrow{\alpha}_k\}
=\{\sum\limits_{i=1}^k{\lambda_i\overrightarrow{\alpha}_i}:
\lambda_1,\lambda_2,\dots,\lambda_k \in \mathbb{Z}_d\}.
\end{array} \end{equation}
$\overrightarrow{\alpha}_1,\overrightarrow{\alpha}_2,\dots,\overrightarrow{\alpha}_k$
are said to be linearly independent if for any $a_1,a_2,\dots,a_k
\in \mathbb{Z}_d$,
$\sum_{i=1}^k{a_i\overrightarrow{\alpha}_i}=\overrightarrow{0}$ if
and only if $a_1=a_2=\dots=a_k=0$. In particular, any $n$ linearly
independent vectors
$\overrightarrow{\alpha}_1,\overrightarrow{\alpha}_2,\dots,\overrightarrow{\alpha}_n$
in $\mathbb{Z}^n_d$ are called a basis of $\mathbb{Z}^n_d$.

Now let us reformulate our problem precisely. Suppose
$g_1,g_2,\dots,g_n$ are independent commuting operators in
$G'^{(d)}_n$ and $S=\langle g_1,g_2,\dots,g_n\rangle$ is a
complete stabilizer. $|\psi_S\rangle$ is the state stabilized by
$S$. Assume that $(T_1,T_2,\dots,T_{m+1})$ is a partition of
$[1,n]$. $A_1,A_2,\dots,A_{m+1}$ are distant parties and $A_i$
holds the subsystem $T_i$ of $|\psi_S\rangle$, for
$i=1,2,\dots,m+1$. Now suppose $A_i$ performs one out of $b_i$
different quantum operations on the subsystem $T_i$, for
$i=1,2,\dots,m$. Then $A_1,A_2,\dots,A_{m}$ send all their
subsystems to $A_{m+1}$. If $A_{m+1}$ can perfectly discriminate
among all possible states, then $(b_1,b_2,\dots,b_m)$ is said to
be an achievable alphabet size for $|\psi\rangle$ with respect to
$(T_1,T_2,\dots,T_{m+1})$. Our primary goal is to determine
whether there exists an achievable alphabet size
$(b_1,b_2,\dots,b_m)$ such that $b_i>d^{|T_i|}$ for at least one
$1 \le i \le m$ and $b_j \ge d^{|T_j|}$ for other $j \neq i$. If
so, $|\psi_S\rangle$ is useful for deterministic distributed dense
coding with respect to $(T_1,T_2,\dots,T_{m+1})$. Our ultimate
goal is to completely determine the set of achievable alphabet
sizes for $|\psi_S\rangle$ with respect to an arbitrary partition
plan.

Now suppose a deterministic distributed dense coding protocol
achieves the alphabet size $(b_1,b_2,\dots,b_m)$ for
$|\psi_S\rangle$ with respect to $(T_1,T_2,\dots,T_{m+1})$ by
setting $A_i$'s encoding operations to be unitary operations
$\{U_{ij}:j=1,2,\dots,b_i\}$, $\forall i =1,2,\dots,m$. Then there
are totally $\prod\limits_{i=1}^{m}{b_i}$ possible encoded states
which are given by
\begin{equation}\begin{array}{l}
|\psi(\overrightarrow{j})\rangle=(U_{1j_1}\otimes U_{2j_2} \otimes \dots \otimes U_{mj_m} \otimes I)|\psi_S\rangle,\\
\label{equ:rhoj}
\end{array}\end{equation}
where $\overrightarrow{j}=(j_1,j_2,\dots,j_m)$ with $1 \le j_i \le
b_i$, $\forall i=1,2,\dots,m$. These states can be perfectly
discriminated by the receiver $A_{m+1}$ if and only if they are
mutually orthogonal, i.e. $\langle
\psi(\overrightarrow{j})|\psi(\overrightarrow{j}')\rangle=0$,
$\forall \overrightarrow{j}\neq \overrightarrow{j}'$. Furthermore,
we can prove that $(b_1,b_2,\dots,b_m)$ satisfies two constraints.
The first one is
\begin{equation}
\prod\limits_{i=1}^{m}{b_i} \le d^{n}, \label{equ:prodbi}
\end{equation}
since there could be at most $d^n$ mutually orthogonal $n$-qudit
states. Apparently the protocol reaches the best efficiency if and
only if $\prod_{i=1}^{m}{b_i}=d^{n}$. In this case, we say that
$|\psi_S\rangle$ is \textit{optimally} useful for deterministic
distributed dense coding with respect to
$(T_1,T_2,\dots,T_{m+1})$. The second constraint is
\begin{equation}\begin{array}{ll}
b_i \le d^{2|T_i|}, & \forall i=1,2,\dots,m.
\end{array}\end{equation}
To see this, one needs to realize that the states
\begin{equation}\begin{array}{ll}
|\phi({j})\rangle&=(U_{1j}\otimes U_{21} \otimes U_{31} \otimes
\dots \otimes U_{m1} \otimes I)
|\psi_S\rangle\\
&=(U_{1j}\otimes I)|\psi'_S\rangle
\end{array}\end{equation}
for $j=1,2,\dots,b_1$ are mutually orthogonal, where
\begin{equation}\begin{array}{l}
|\psi'_S\rangle=(I \otimes U_{21} \otimes U_{31} \otimes \dots
\otimes U_{m1} \otimes I) |\psi_S\rangle.
\end{array}\end{equation}
This means that $b_1$ is an achievable alphabet size for
$|\psi'_S\rangle$ with respect to the bipartition
$(T_1,\bigcup_{i=2}^{m+1}{T_i})$. Since in a bipartite dense
coding scheme with an arbitrary $d_1 \times d_2$ state (where the
$d_1$-dimensional subsystem is held by the sender) the alphabet
size that cannot exceed $d^2_1$ \cite{BD04}, we obtain $b_1 \le
d^{2|T_1|}$. Similarly, $b_i \le d^{2|T_i|}$, $\forall
i=2,3,\dots,m$. The second constraint tells us no matter how we
group the $n$ qudits and how we encode, eventually every sender
can acquire at most twice the classical information capacity of
the original noiseless quantum channel.

From now on we will focus on deterministic distributed dense
coding schemes whose encoding operations are chosen from the
generalized Pauli group on multiple qudits. Let us first look at
the effect of this kind of operations on the state
$|\psi_S\rangle$. From the fact that $|\psi_S\rangle$ is
stabilized by $S=\langle g_1,g_2,\dots,g_n\rangle$, we know for
any $g \in G_n$, $g|\psi_S\rangle$ is the state stabilized by
\begin{equation}\begin{array}{ll}
gSg^{\dagger}&=\langle gg_1g^{\dagger},gg_2g^{\dagger},\dots,gg_ng^{\dagger}\rangle\\
&=\langle \omega^{\chi(g)\Lambda\chi(g_1)}g_1,
\omega^{\chi(g)\Lambda\chi(g_2)}g_2,
\dots,\omega^{\chi(g)\Lambda\chi(g_n)}g_n\rangle, \label{equ:gsg}
\end{array}\end{equation}
where the second equality comes from Eq.(\ref{equ:ghg}). In other
words, $g|\psi_S\rangle$ becomes the simultaneous eigenstate of
$g_1,g_2,\dots,g_n$ with the eigenvalues
$\omega^{-\chi(g)\Lambda\chi(g_1)},
\omega^{-\chi(g)\Lambda\chi(g_2)},
\dots,\omega^{-\chi(g)\Lambda\chi(g_n)}$ respectively. Now we
introduce a map $\Gamma$ as follows: if $|\psi\rangle$ is the
simultaneous eigenstate of $g_1, g_2, \dots, g_n$ corresponding to
the eigenvalues $\omega^{x_1}, \omega^{x_2}, \dots, \omega^{x_n}$
for some $x_1,x_2,\dots,x_n \in \mathbb{Z}_d$, then
$\Gamma(|\psi\rangle)=(x_1,x_2,\dots,x_n)^T$; otherwise,
$\Gamma(|\psi\rangle)$ is not defined. So for any $g \in
G^{(d)}_n$, we have
\begin{equation}\begin{array}{l}
\Gamma(g|\psi_S\rangle)\\
=(-\chi(g)\Lambda\chi(g_1), -\chi(g)\Lambda\chi(g_2),
\dots,-\chi(g)\Lambda\chi(g_n))\\
=(\sum\limits_{j=1}^{n}{a_jb_{1j}}-\sum\limits_{j=1}^{n}{b_ja_{1j}},
\sum\limits_{j=1}^{n}{a_jb_{2j}}-\sum\limits_{j=1}^{n}{b_ja_{2j}},\\
\dots,
\sum\limits_{j=1}^{n}{a_jb_{nj}}-\sum\limits_{j=1}^{n}{b_ja_{nj}})^T\\
=\sum\limits_{j=1}^{n}{a_j(b_{1j},b_{2j},\dots,b_{nj})^{T}}
-\sum\limits_{j=1}^{n}{b_j(a_{1j},a_{2j},\dots,a_{nj})^{T}}\\
=\sum\limits_{j=1}^{n}{a_j\overrightarrow{\beta_j}}
-\sum\limits_{j=1}^{n}{b_j\overrightarrow{\alpha_j}}\\
\label{equ:grhog}
\end{array}\end{equation}
where we suppose $\chi(g)=(a_1,a_2,\dots,a_n,b_1,b_2,\dots,b_n)$,
$\chi(g_i)=(a_{i1}, a_{i2}, \dots, a_{in}, b_{i1}, b_{i2}, \dots,
b_{in})$, $\forall i =1,2,\dots,n$, and
\begin{equation}\begin{array}{l}
\overrightarrow{\alpha_j}=(a_{1j},a_{2j},\dots,a_{nj})^{T},\\
\overrightarrow{\beta_j}=(b_{1j},b_{2j},\dots,b_{nj})^{T},\\
\end{array}\end{equation}
$\forall j =1,2,\dots,n$. Note that $\overrightarrow{\alpha_j}$
and $\overrightarrow{\beta_j}$ are exactly the $j$-th and
$(j+n)$-th columns of the check matrix $M$ for $g_1,g_2,\dots,g_n$
respectively, $\forall j =1,2,\dots,n$.

Now suppose we have a valid deterministic dense coding protocol in
which $A_i$'s encoding operations are
\begin{equation}\begin{array}{l}
U_{ij}=\bigotimes\limits_{l \in T_i}{\sigma_{a_{ijl},b_{ijl}}}
\end{array}\end{equation}
for $j=1,2,\dots,b_i$, where $a_{ijl},b_{ijl} \in \mathbb{Z}_d$,
$\forall i =1,2,\dots,m$. Then by Eqs.(\ref{equ:rhoj}) and
(\ref{equ:grhog}),
\begin{equation}\begin{array}{l}
\Gamma(|\psi(\overrightarrow{j})\rangle)
=\sum\limits_{i=1}^{m}\sum\limits_{l \in
T_i}({a_{ij_il}\overrightarrow{\beta}_l-b_{ij_il}\overrightarrow{\alpha}_l})
=\sum\limits_{i=1}^{m}{\overrightarrow{\gamma}_{ij_i}},
\end{array}\end{equation}
where
\begin{equation}\begin{array}{ll}
\overrightarrow{\gamma}_{ij_i}
&=\sum\limits_{l \in T_i}{a_{ij_il}\overrightarrow{\beta}_l-b_{ij_il}\overrightarrow{\alpha}_l}\\
&\in S_i \equiv span\{\overrightarrow{\alpha}_l,
\overrightarrow{\beta}_l: l \in T_i\}. \label{equ:definegamma}
\end{array}\end{equation}
So for any $\overrightarrow{j} \neq \overrightarrow{j'}$,
$|\psi(\overrightarrow{j})\rangle$ and
$|\psi(\overrightarrow{j}')\rangle$ are orthogonal if and only if
$\Gamma(|\psi(\overrightarrow{j})\rangle) \neq
\Gamma(|\psi(\overrightarrow{j}')\rangle)$. Therefore we have
\begin{equation}\begin{array}{ll}
\sum\limits_{i=1}^{m}{\overrightarrow{\gamma}_{ij_i}} \neq
\sum\limits_{i=1}^{m}{\overrightarrow{\gamma}_{ij'_i}},& \forall
\overrightarrow{j} \neq \overrightarrow{j}'. \label{equ:gammaij}
\end{array}\end{equation}

Conversely, suppose we are given a set of vectors
$\{\overrightarrow{\gamma}_{ij} \in S_i: i=1,2,\dots,m, j
=1,2,\dots,b_i\}$ which satisfy inequality (\ref{equ:gammaij}).
Since $S_i=span\{\overrightarrow{\alpha}_l,
\overrightarrow{\beta}_l: l \in T_i\}$,
$\overrightarrow{\gamma}_{ij}$ can be written as
\begin{equation}\begin{array}{l}
\overrightarrow{\gamma}_{ij}=\sum\limits_{l \in
T_i}{a_{ijl}\overrightarrow{\beta}_l-b_{ijl}\overrightarrow{\alpha}_l}
\end{array}\end{equation}
for some $a_{ijl},b_{ijl} \in \mathbb{Z}_d$. Then consider the
protocol in which $A_i$ uses the encoding operations
$\{U_{ij}=\bigotimes_{l \in
T_i}{\sigma_{a_{ijl},b_{ijl}}}:j=1,2,\dots,b_i\}$, $\forall i
=1,2,\dots,m$. One can easily see that it is also a valid
deterministic dense coding protocol.

Summarizing the argument in the above two paragraphs, we know
there exists a protocol which achieves the alphabet size
$(b_1,b_2,\dots,b_m)$ by using generalized Pauli group elements to
encode if and only if there exist vectors
$\{\overrightarrow{\gamma}_{ij} \in S_i: i =1,2,\dots,m, j
=1,2,\dots,b_i\}$ which satisfy inequality (\ref{equ:gammaij}).
Thus our problem can be rephrased as follows: given the subspaces
$S_1,S_2,\dots,S_m$ of $\mathbb{Z}^n_d$, for any
$(b_1,b_2,\dots,b_m) \in \mathbb{Z}^m$, do there exist vectors
$\{\overrightarrow{\gamma}_{ij} \in S_i:i =1,2,\dots,m, j =1,2,\dots, b_i\}$
satisfying inequality (\ref{equ:gammaij})?

One can easily see that a necessary condition for
$\sum\limits_{i=1}^{m}{\overrightarrow{\gamma}_{ij_i}}=\sum\limits_{i=1}^{m}{\overrightarrow{\gamma}_{ij'_i}}$
is the vectors
$\{\overrightarrow{\gamma}_{ij_i}-\overrightarrow{\gamma}_{ij'_i}:
i =1,2,\dots,m\}$ are linearly dependent. With this observation,
we obtain a sufficient condition for $|\psi_S\rangle$ to be useful
for deterministic distributed dense coding with respect to
$(T_1,T_2,\dots,T_{m+1})$, as the following theorem states:

\begin{theorem}
If there exist $R_i,Q_i \subset T_i$, $\forall i =1,2,\dots,m$,
such that: (1) $|R_i|+|Q_i| > |T_i|$ for at least one $1 \le i \le
m$, while $|R_j|+|Q_j| \ge |T_j|$ for other $j \neq i$; (2) the
vectors $\{\overrightarrow{\alpha}_k: k \in \cup_{i=1}^{m}{R_i}\}
\cup \{\overrightarrow{\beta}_l: l \in \cup_{i=1}^{m}{Q_i}\}$ are
linearly independent, then $|\psi_S\rangle$ is useful for
deterministic distributed dense coding with respect to
$(T_1,T_2,\dots,T_{m+1})$.
\end{theorem}
\textit{Proof:} Consider the following protocol: $\forall i
=1,2,\dots,m$, $A_i$'s encoding operations are
\begin{equation}\begin{array}{l}
U_i(\{a_{ik},b_{ik}\}_{k \in T_i})=\bigotimes\limits_{k \in T_i}{\sigma_{a_{ik},b_{ik}}},\\
\end{array}\end{equation}
where $a_{ik}=0,1,\dots,d-1$, for $k \in Q_i$; $a_{ik}=0$, for $k
\in T_i-Q_i$; $b_{ik}=0,1,\dots,d-1$, for $k \in R_i$; $b_{ik}=0$,
for $k \in T_i-R_i$. There are totally $d^{|R_i|+|Q_i|}$ different
choices of $\{a_{ik},b_{ik}\}_{k \in T_i}$. In other words, the
alphabet size of $A_i$ is $d^{|R_i|+|Q_i|}$. If we prove that this
protocol is valid, then by condition (1), $|\psi_S\rangle$ is
useful for deterministic distributed dense coding with respect to
$(T_1,T_2,\dots,T_{m+1})$.

By Eq.(\ref{equ:definegamma}), the vector corresponding to
$U_i(\{a_{ik},b_{ik}\}_{k \in T_i})$ is
\begin{equation}\begin{array}{ll}
\overrightarrow{\gamma}_i(\{a_{ik},b_{ik}\}_{k \in T_i})
&=\sum\limits_{k \in T_i}{a_{ik}\overrightarrow{\beta}_k}
-\sum\limits_{k \in T_i}{b_{ik}\overrightarrow{\alpha}_k}\\
&=\sum\limits_{k \in Q_i}{a_{ik}\overrightarrow{\beta}_k}
-\sum\limits_{k \in R_i}{b_{ik}\overrightarrow{\alpha}_k}.
\label{equ:gammaab}
\end{array}\end{equation}

Now suppose for some $\{a_{ik},b_{ik}\}$, $\{a'_{ik},b'_{ik}\}$,
we have
\begin{equation}\begin{array}{l}
\sum\limits_{i=1}^m{\overrightarrow{\gamma}_i(\{a_{ik},b_{ik}\}_{k
\in T_i})}
=\sum\limits_{i=1}^m{\overrightarrow{\gamma}_i(\{a'_{ik},b'_{ik}\}_{k \in T_i})}.\\
\end{array}\end{equation}
Then by Eq.(\ref{equ:gammaab}), we obtain
\begin{equation}\begin{array}{l}
\sum\limits_{i=1}^m(\sum\limits_{k \in
Q_i}{(a_{ik}-a'_{ik})\overrightarrow{\beta}_k}
-\sum\limits_{k \in R_i}{(b_{ik}-b'_{ik})\overrightarrow{\alpha}_k})=\overrightarrow{0}.\\
\end{array}\end{equation}
Since $\{\overrightarrow{\alpha}_k: k \in \cup_{i=1}^{m}{R_i}\}
\cup \{\overrightarrow{\beta}_k: k \in \cup_{i=1}^{m}{Q_i}\}$ are
linearly independent, this equation implies $a_{ik}=a'_{ik}$,
$\forall k \in Q_i$ and $b_{ik}=b'_{ik}$, $\forall k \in R_i$,
$\forall i=1,2,\dots,m$. So the vectors
$\{\overrightarrow{\gamma}_i(\{a_{ik},b_{ik}\}_{k \in T_i})\}$
satisfy inequality (\ref{equ:gammaij}). In other words, this
protocol is valid. This ends the proof \hfill $\blacksquare$

With the help of this theorem, we find that when $d$ is prime, the
power of $n$-qudit stabilizer states in deterministic distributed
dense coding is strong, as the following corollary states:

\begin{corollary}
If $d$ is prime, then any genuinely entangled $n$-qudit stabilizer
state is optimally useful for deterministic distributed dense
coding with respect to at least one partition of $[1,n]$.
\end{corollary}
\textit{Proof:} In Ref.\cite{NC00}, the authors present a
procedure which can transform any $k \times 2n$ check matrix for
$S=\langle g_1,g_2,\dots,g_k\rangle$ (where $d=2$) into the
following standard form
\begin{equation}\begin{array}{l}
M=\begin{pmatrix}
I_r & A_1 & A_2 & | & B & 0 & C\\
0 & 0 & 0 & | & D & I_{k-r} & E\\
\end{pmatrix},
\end{array}\end{equation}
(re-labelling the original $n$ qudits and re-selecting stabilizer
generators if necessary), where $A_1,A_2, B,C,D,E$ are matrices of
size $r \times (k-r)$, $r \times (n-k)$, $r \times r$, $r \times
(n-k)$, $(k-r) \times r$, $(k-r) \times (n-k)$ respectively, for
some $r \le k$. Their procedure includes three basic kinds of
operations about the original matrix: swapping rows, swapping
columns and adding one row to another. We realize that their
conclusion can be readily extended to arbitrary prime dimensions,
since the essential prerequisite of their method is that
$\mathbb{Z}_d$ needs to be a field.

Now $|\psi_S\rangle$ is stabilized by a complete stabilizer
$S=\langle g_1,g_2,\dots,g_n\rangle$. In this case, the above
standard form reduces into
\begin{equation}\begin{array}{l}
M=\begin{pmatrix}
I_r & A_1 & | & B & 0 \\
0 & 0 & | & D & I_{n-r}\\
\end{pmatrix},
\end{array}
\label{equ:standard}
\end{equation}
where $A_1,B,D$ are matrices of size $r \times (n-r)$, $r \times
r$, $(n-r) \times r$ respectively, for some $r \le n$. Then we
have
\begin{equation}\begin{array}{ll}
0&=M\Lambda_{n} M^{\dagger}\\
&=\begin{pmatrix}
B-B^T  & -A_1-D^T \\
A^T_1+D & 0 \\
\end{pmatrix},
\end{array}\end{equation}
which yields $B=B^T$ and $A_1+D^T=0$.

Now we prove $D \neq 0$ by contradiction. Assume $D=0$. Define
\begin{equation}\begin{array}{ll}
M_1=\begin{pmatrix}
I_r  & | & B \\
0 & | & D \\
\end{pmatrix}.
\end{array}
\end{equation}
Then we have
\begin{equation}\begin{array}{l}
M_1\Lambda_{r}M^{\dagger}_1 =\begin{pmatrix}
B-B^T  & -D^T \\
D  & 0 \\
\end{pmatrix}
=0.
\end{array}\end{equation}
For any $T \subset [1,n]$ and any $g=\gamma^c \sigma_{a_1,b_1}
\otimes \sigma_{a_2,b_2} \otimes \dots \otimes \sigma_{a_n,b_n}
\in G^{(d)}_n$, define the restriction of $g$ on $T$ as
\begin{equation}\begin{array}{l}
g^{(T)}=\bigotimes\limits_{k \in T}{\sigma_{a_k,b_k}}.
\end{array}\end{equation}
Then one can see that $M_1\Lambda_{r}M^{\dagger}_1=0$ implies
$g'^{([1,r])}_{1},g'^{([1,r])}_{2},\dots,g'^{([1,r])}_{n}$
mutually commute, where $g'_i$ is the stabilizer generator
corresponding to the $i$-th row of $M$ in Eq.(\ref{equ:standard}),
for $i=1,2,\dots,n$. Thus by lemma 1 of Ref. \cite{WY07},
$|\psi_S\rangle$ should be separable with respect to the
bipartition $([1,r],[r+1,n])$. This contradicts with the given
fact that $|\psi_S\rangle$ is genuinely entangled. So $D \neq 0$.

Now suppose the entry on the $k$-th row and $l$-th column of $D$
is nonzero. Assume the $l$-th column of $B$ is
$(b_1,b_2,\dots,b_r)^T$ and the $l$-th column of $D$ is
$(d_1,d_2,\dots,d_{n-r})^T$ with $d_{k} \neq 0$. Then the
$(n+l)$-th column of $M$ is
$\overrightarrow{\beta}_l=(b_1,b_2,\dots,b_r,d_1,d_2,\dots,d_{n-r})^T$.
Note that the $i$-th column of $M$ is
$\overrightarrow{\alpha}_i=(0,\dots,0,1,0,\dots,0)^T$ where $1$ is
the $i$-th element, for $i=1,2,\dots,r$. Also, the $(n+i)$-th
column of $M$ is
$\overrightarrow{\beta}_i=(0,\dots,0,1,0,\dots,0)^T$ where $1$ is
the $i$-th element, for $i=r+1,r+2,\dots,n$.

Consider the partition $(T_1,T_2,\dots,T_n)$ with $T_i=\{i\}$,
$\forall i=1,2,\dots,r+k-1$; $T_i=\{i+1\}$, $\forall
i=r+k,r+k+1,\dots,n-1$; $T_n=\{r+k\}$. In other words, the
receiver holds the $(r+k)$-th qudit and $n-1$ senders each hold
one of the other $n-1$ qudits. Consider the vectors
$\overrightarrow{\alpha}_1, \overrightarrow{\alpha}_2, \dots,
\overrightarrow{\alpha}_r$, $\overrightarrow{\beta}_l$,
$\overrightarrow{\beta}_{r+1}, \overrightarrow{\beta}_{r+2},
\dots, \overrightarrow{\beta}_{r+k-1},
\overrightarrow{\beta}_{r+k+1}, \dots,
\overrightarrow{\beta}_{n}$. They are linearly independent.
Actually, suppose for $c_1,c_2,\dots,c_{r+k-1},
c_{r+k+1},\dots,c_{n}, \lambda \in \mathbb{Z}_d$, we have
\begin{equation}\begin{array}{ll}
\overrightarrow{0}
&=\sum\limits_{i=1}^{r}{c_i\overrightarrow{\alpha}_i}
+\sum\limits_{i=r+1}^{r+k-1}{c_i\overrightarrow{\beta}_i}
+\sum\limits_{i=r+k+1}^{n}{c_i\overrightarrow{\beta}_i}
+\lambda\overrightarrow{\beta}_l\\
&=(c_1+b_1\lambda,c_2+b_2\lambda,\dots,c_r+b_r\lambda,\\
&c_{r+1}+d_1\lambda, c_{r+2}+d_2\lambda,\dots,
c_{r+k-1}+d_{k-1}\lambda,\\
&d_{k}\lambda,c_{r+k-1}+d_{k+1}\lambda,\dots,
c_{n}+d_{n-r}\lambda)^T.
\end{array}\end{equation}
Since $d_{k}\neq 0$, the entry $d_{k}\lambda=0$ implies
$\lambda=0$. Taking this back to the above equation, we obtain
$c_1=c_2=\dots=c_{r+k-1}=c_{r+k+1}=\dots=c_n=0$. Now define
$R_i=T_i$, $\forall i=1,2,\dots,r$; $R_i=\emptyset$, $\forall
i=r+1,r+2,\dots,n-1$; $Q_i=\emptyset$, $\forall i
=1,2,\dots,l-1,l+1,l+2,\dots,r$; $Q_i=T_i$, $\forall
i=l,r+1,r+2,\dots,n-1$. Then $\{R_i,Q_i: i =1,2,\dots,n-1]\}$
satisfy the conditions of theorem 1. Furthermore, note that
$\sum_{i=1}^{n-1}(|R_i|+|Q_i|)=n$. So by the proof of theorem 1,
we know $|\psi_S\rangle$ is optimally useful for deterministic
distributed dense coding with respect to $(T_1,T_2,\dots,T_n)$.
\hfill $\blacksquare$

\textit{Remark.} From theorem 1 and corollary 1, we see that the
linear independency among the columns of check matrix can affect
the dense coding power of $|\psi\rangle$. The more linearly
independent they are, the more powerful $|\psi_S\rangle$ is for
dense coding.

The dense coding protocol given by the proof of theorem 1 always
has alphabet size of the form $(d^{a_1},d^{a_2},\dots, d^{a_m})$
for some integers $a_1,a_2,\dots,a_m$. One may wonder whether a
wider class of alphabet sizes can be reached. Indeed this is true.
In what follows, we will propose a method to partially solve the
general problem of determining the whole set of achievable
alphabet sizes.

Now suppose $\{\overrightarrow{x}_1,\overrightarrow{x}_2,
\dots,\overrightarrow{x}_n\}$ is an arbitrary basis of
$\mathbb{Z}^{n}_d$. Let $\overrightarrow{X}$ denote this basis.
For $j=1,2,\dots,n$, define
\begin{equation}\begin{array}{l}
W_j=span\{\overrightarrow{x}_{j}, \overrightarrow{x}_{j+1}, \dots,
\overrightarrow{x}_{n}\}.
\end{array}\end{equation}
Also define $W_{n+1}=\emptyset$. For any
$\overrightarrow{z}=\sum_{i=1}^n{\lambda_i\overrightarrow{x}_i}
\in \mathbb{Z}^{n}_d$, define
$C_j(\overrightarrow{z};\overrightarrow{X})=\lambda_j$, $\forall
j=1,2,\dots,n$. For $j=1,2,\dots,n$, define
\begin{equation}
P_j=\{1 \le i \le m: S_i \cap (W_j-W_{j+1}) \neq \emptyset\}.
\label{equ:pj}
\end{equation}
Then for all $i \in P_j$, choose $\overrightarrow{z}_{ij} \in S_i
\cap (W_j-W_{j+1})$. Let
\begin{equation}\begin{array}{l}
c_{ij}=C_j(\overrightarrow{z}_{ij};\overrightarrow{X}).
\end{array}\end{equation}
Note that by the definition of $\overrightarrow{z}_{ij}$, it
satisfies
\begin{equation}\begin{array}{ll}
C_{k}(\overrightarrow{z}_{ij};\overrightarrow{X})=0, & \forall
k<j. \label{equ:cjj}
\end{array}\end{equation}

For any $t\ge 1$, define
\begin{equation}\begin{array}{ll}
A(P_j;t)&=\{(a_1,a_2,\dots,a_m) \in \mathbb{Z}^{m}: \forall i
\notin P_j,
a_i=1;\\
&\prod\limits_{i=1}^{m}{a_i} \le t\}.
\end{array}\end{equation}

For $j=1,2,\dots,n$, choose
$\overrightarrow{a}_j=(a_{j1},a_{j2},\dots,a_{jm}) \in A(P_j;d)$.
Then define
\begin{equation}\begin{array}{llll}
Q_j=\{i \in P_j: \exists \eta \in \mathbb{Z}_d, &s.t.&
\eta c_{ij} \equiv \prod\limits_{t=1}^{i-1}{a_{jt}} (mod & d)\},\\
\end{array}\end{equation}
and furthermore,
\begin{equation}\begin{array}{l}
\overrightarrow{b}_j=F(\overrightarrow{a}_j;Q_j)=(b_{j1},b_{j2},\dots,b_{jm}),
\end{array}\end{equation}
where $b_{ji}=a_{ji}$, $\forall i \in Q_j$; $b_{ji}=1$, $\forall i
\not\in Q_j$. Let $B(\overrightarrow{X})$ denote the set of
$(\overrightarrow{b}_1,\overrightarrow{b}_2,
\dots,\overrightarrow{b}_n)$ that can be obtained by this
procedure.

With these definitions and notations introduced above, we have the
following theorem:

\begin{theorem}
For any $(\overrightarrow{b}_1,\overrightarrow{b}_2,
\dots,\overrightarrow{b}_n) \in B(\overrightarrow{X})$,
$(\prod\limits_{j=1}^{n}{b_{j1}}, \prod\limits_{j=1}^{n}{b_{j2}},
\dots, \prod\limits_{j=1}^{n}{b_{jm}})$ is an achievable alphabet
size for $|\psi_S\rangle$ with respect to
$(T_1,T_2,\dots,T_{m+1})$.
\end{theorem}

\textit{Proof:} In what follows, if not explicitly pointed out,
all computations will be taken over $\mathbb{Z}_d$. By definition,
$\forall j=1,2,\dots,n$, $\forall i \in Q_j$, there exists
$\eta_{ij} \in \mathbb{Z}_d$ such that
\begin{equation}\begin{array}{ll}
\eta_{ij} c_{ij}=\prod\limits_{t=1}^{i-1}{a_{jt}}.
\end{array}\end{equation}
Define
\begin{equation}\begin{array}{l}
\overrightarrow{y}_{ij}=\eta_{ij} \overrightarrow{z}_{ij} \in S_i.
\end{array}\end{equation}
Then we have
\begin{equation}\begin{array}{l}
C_j(\overrightarrow{y}_{ij};\overrightarrow{X})
=\prod\limits_{t=1}^{i-1}{a_{jt}}. \label{equ:cjy}
\end{array}\end{equation}
Moreover, by Eq.(\ref{equ:cjj}),
\begin{equation}\begin{array}{ll}
C_k(\overrightarrow{y}_{ij};\overrightarrow{X})=0,& \forall k<j.
\label{equ:cky}
\end{array}\end{equation}
For all $i \not\in Q_j$, define
\begin{equation}\begin{array}{ll}
\overrightarrow{y}_{ij}=\overrightarrow{0}.
\end{array}\end{equation}

Now for $i=1,2,\dots,m$, we define
\begin{equation}
\overrightarrow{\gamma}(i;\overrightarrow{\lambda}_i)=
\sum\limits_{j=1}^{n}{\lambda_{ij}\overrightarrow{y}_{ij}} \in
S_i, \label{equ:gammilambdai}
\end{equation}
where
$\overrightarrow{\lambda}_i=(\lambda_{i1},\lambda_{i2},\dots,\lambda_{in})$
with $1 \le \lambda_{ij} \le b_{ji}$, $\forall j=1,2,\dots,n$. We
will prove the vectors
$\{\overrightarrow{\gamma}(i;\overrightarrow{\lambda}_i)\}$
satisfy inequality (\ref{equ:gammaij}).

Suppose for some $\{\overrightarrow{\lambda}_{i}\}$,
$\{\overrightarrow{\mu}_i\}$,
\begin{equation}
\sum\limits_{i=1}^{m}{\overrightarrow{\gamma}(i;\overrightarrow{\lambda}_i)}
=\sum\limits_{i=1}^{m}{\overrightarrow{\gamma}(i;\overrightarrow{\mu}_i)},
\end{equation}
or equivalently,
\begin{equation}
\sum\limits_{i=1}^{m}\sum\limits_{j=1}^{n}{({\lambda}_{ij}-{\mu}_{ij})\overrightarrow{y}_{ij}}
=0. \label{equ:lqm}
\end{equation}

Note that for $j=1,2,\dots,n$, $\forall i \not\in Q_j$,
$\lambda_{ij}=\mu_{ij}=1$ because by definition $1 \le
\lambda_{ij},\mu_{ij} \le b_{ji}=1$. So Eq.(\ref{equ:lqm}) reduces
into
\begin{equation}
\sum\limits_{j=1}^{n}\sum\limits_{i \in
Q_j}{({\lambda}_{ij}-{\mu}_{ij})\overrightarrow{y}_{ij}}=0.
\label{equ:lm}
\end{equation}

If we write the left-hand side of Eq.(\ref{equ:lm}) as linear
combination of the basis
$\overrightarrow{x}_1,\overrightarrow{x}_2,\dots,\overrightarrow{x}_n$,
then the coefficient corresponding to $\overrightarrow{x}_1$
should be zero, i.e.
\begin{equation}\begin{array}{l}
0=\sum\limits_{j=1}^{n}\sum\limits_{i \in Q_j}{
({\lambda}_{ij}-{\mu}_{ij})C_1(\overrightarrow{y}_{ij};\overrightarrow{X})}\\
=\sum\limits_{i \in Q_1}{({\lambda}_{i1}-{\mu}_{i1})\prod\limits_{t=1}^{i-1}{a_{1t}}},\\
\end{array}\end{equation}
where the second equality comes from Eqs.(\ref{equ:cjy}) and
(\ref{equ:cky}).

Now define
\begin{equation}
R_1=\{i \in Q_1: a_{1i} \neq 1\}.
\end{equation}
Then $\forall i \in Q_1-R_1$, $b_{1i}=a_{1i}=1$, and consequently
$\lambda_{i1}=\mu_{i1}=1$ since by definition $1 \le \lambda_{i1},
\mu_{i1} \le b_{1i}$. So
\begin{equation}\begin{array}{l}
0=\sum\limits_{i \in
Q_1}{({\lambda}_{i1}-{\mu}_{i1})\prod\limits_{t=1}^{i-1}{a_{1t}}}
=\sum\limits_{i \in
R_1}{({\lambda}_{i1}-{\mu}_{i1})\prod\limits_{t=1}^{i-1}{a_{1t}}}.
\label{equ:zero}
\end{array}\end{equation}

Note that the above additions and multiplications are taken over
$\mathbb{Z}_d$. Eq.(\ref{equ:zero}) actually means
\begin{equation}\begin{array}{ll}
\sum\limits_{i \in
R_1}{({\lambda}_{i1}-{\mu}_{i1})\prod\limits_{t=1}^{i-1}{a_{1t}}}\equiv
0(mod & d).
\end{array}\end{equation}
Now we turn back to normal computation over $\mathbb{Z}$. We
actually can prove
\begin{equation}\begin{array}{l}
\sum\limits_{i \in
R_1}{({\lambda}_{i1}-{\mu}_{i1})\prod\limits_{t=1}^{i-1}{a_{1t}}}=0.
\end{array}\end{equation}
To see this, one only needs to realize
\begin{equation}\begin{array}{l}
|\sum\limits_{i \in R_1}{({\lambda}_{i1}-{\mu}_{i1})\prod\limits_{t=1}^{i-1}{a_{1t}}}|\\
\le \sum\limits_{i \in R_1}{(b_{1i}-1)\prod\limits_{t=1}^{i-1}{a_{1t}}}\\
=\sum\limits_{i \in R_1}{(\prod\limits_{t=1}^{i}{a_{1t}}-\prod\limits_{t=1}^{i-1}{a_{1t}})}\\
\le \sum\limits_{i=1}^{m}{(\prod\limits_{t=1}^{i}{a_{1t}}-\prod\limits_{t=1}^{i-1}{a_{1t}})}\\
=\prod\limits_{t=1}^{m}{a_{1t}}-1\\
\le d-1,
\end{array}\end{equation}
where the first inequality comes from $1 \le \lambda_{i1},
\mu_{i1} \le b_{1i}$, the second equality comes from
$b_{1i}=a_{1i}$, $\forall i \in R_1 \subset Q_1$, and the last
inequality comes from $(a_{11},a_{12},\dots,a_{1m}) \in A(P_1;d)$.

Suppose $i_1$, $i_2$ is the smallest and second smallest number in
$R_1$. Then for $\forall i \ge i_2 \in R_1$,
$\prod\limits_{t=1}^{i-1}{a_{1t}}$ is a multiple of
$\prod\limits_{t=1}^{i_2-1}{a_{1t}}$. Consequently, from
$\sum\limits_{i \in
R_1}{({\lambda}_{i1}-{\mu}_{i1})\prod\limits_{t=1}^{i-1}{a_{1t}}}=0$
we get that
${({\lambda}_{i_11}-{\mu}_{i_11})\prod\limits_{t=1}^{i_1-1}{a_{1t}}}$
is a multiple of $\prod\limits_{t=1}^{i_2-1}{a_{1t}}$. But on the
other hand,
\begin{equation}\begin{array}{l}
|{({\lambda}_{i_11}-{\mu}_{i_11})\prod\limits_{t=1}^{i_1-1}{a_{1t}}}|
\le (b_{1i_1}-1) \prod\limits_{t=1}^{i_1-1}{a_{1t}}\\
=(a_{1i_1}-1) \prod\limits_{t=1}^{i_1-1}{a_{1t}} <
\prod\limits_{t=1}^{i_1}{a_{1t}}
\le \prod\limits_{t=1}^{i_2-1}{a_{1t}}.\\
\end{array}\end{equation}
So we must have ${\lambda}_{i_11}={\mu}_{i_11}$, which furthermore
implies
\begin{equation}\begin{array}{l}
0=\sum\limits_{i \in
R_1-\{i_1\}}{({\lambda}_{i1}-{\mu}_{i1})\prod\limits_{t=1}^{i-1}{a_{1t}}}.
\end{array}\end{equation}
Repeating the above argument for $i_2$ and the third smallest
number $i_3$ in $R_1$, one can get
${\lambda}_{i_21}={\mu}_{i_21}$. So by iterating this procedure
one can eventually get ${\lambda}_{i1}={\mu}_{i1}$, $\forall i \in
R_1$.

Summarizing the above argument, we obtain
${\lambda}_{i1}={\mu}_{i1}$, $\forall i =1,2,\dots,m$. Taking this
back to Eq.(\ref{equ:lm}), we get
\begin{equation}
\sum\limits_{j=2}^{n}\sum\limits_{i \in Q_j}{
({\lambda}_{ij}-{\mu}_{ij})\overrightarrow{y}_{ij}} =0.
\label{equ:lm2}
\end{equation}
Again, by taking a similar analysis for $\overrightarrow{x}_2$, we
can obtain ${\lambda}_{i2}={\mu}_{i2}$, $\forall i =1,2,\dots,m$.

Repeat this procedure, and eventually we prove
${\lambda}_{ij}={\mu}_{ij}$, $\forall i=1,2,\dots,m$, $\forall
j=1,2,\dots,n$. Therefore
$\{\overrightarrow{\gamma}(i;\overrightarrow{\lambda}_i) \in S_i:
1 \le i \le m\}$ satisfy inequality (\ref{equ:gammaij}), where
$\overrightarrow{\lambda}_i=(\lambda_{i1},\lambda_{i2},\dots,\lambda_{in})$
with $1 \le \lambda_{ij} \le b_{ji}$, $\forall j=1,2,\dots,n$. So
$(\prod\limits_{j=1}^{n}{b_{j1}}, \prod\limits_{j=1}^{n}{b_{j2}},
\dots, \prod\limits_{j=1}^{n}{b_{jm}})$ is an achievable alphabet
size for $|\psi_S\rangle$ with respect to
$(T_1,T_2,\dots,T_{m+1})$. \hfill $\blacksquare$

\textit{Remark 1.} One can see that several ingredients of theorem
2 can be chosen freely. These ingredients include the basis
$\overrightarrow{x}_i$, the vectors $\overrightarrow{y}_{ij} \in
S_i$ and $(a_{j1},a_{j2},\dots,a_{jm}) \in A(P_j;d)$. Every
possible selection of these variables can lead to an achievable
alphabet size by applying theorem 2.

\textit{Remark 2.} One can see that the overall alphabet size of
all senders is
\begin{equation}\begin{array}{l}
\prod\limits_{i=1}^{m}\prod\limits_{j=1}^{n}{b_{ji}}
=\prod\limits_{j=1}^{n}\prod\limits_{i \in Q_j}{a_{ji}}\\
\le \prod\limits_{j=1}^{n}\prod\limits_{i \in P_j}{a_{ji}}
=\prod\limits_{j=1}^{n}\prod\limits_{i=1}^{m}{a_{ji}}\\
\le d^n, \label{equ:global}
\end{array}\end{equation}
where the first equality comes from the definition of $b_{ji}$,
the second inequality comes from $Q_j \subset P_j$ and $a_{ji} \ge
1$, the third inequality comes from $a_{ji}=1$, $\forall i \not\in
P_j$, and the last inequality comes from
$(a_{j1},a_{j2},\dots,a_{jm}) \in A(P_j;d)$. So as long as
$Q_j=P_j$ and $\prod\limits_{i=1}^{m}{a_{ji}}=d$, $\forall
j=1,2,\dots,n$, the alphabet size obtained by theorem 2 is
optimal.

\section{Illustrations}

In this section we will analyze several states by applying our
theorems. In each example, the matrices $X$, $Z$ are
$X^{(d)},Z^{(d)}$ defined by Eq.(\ref{equ:pauli}) with the
corresponding dimension $d$, and similarly for $\sigma_{i,j}$. We
will also use the notation $X_j$ to denote the operation $X$
acting on the $j$th qudit and similarly for $Z_j$. Moreover, all
the entries of check matrices range over $\mathbb{Z}_d$. So we can
use $-c$ to equivalently denote $d-c$, $\forall c=1,2,\dots,d-1$.

We will consider four examples. The first two examples are
re-examinations of old results from our perspective. The third and
fourth examples are detailed illustrations of how to utilize
theorem 1 and theorem 2 respectively.

\begin{example}
Let us begin with the standard bipartite dense coding protocol.
Let
\begin{equation}\begin{array}{l}
|\Phi^{+}\rangle=\frac{1}{\sqrt{d}}\sum\limits_{i=0}^{d-1}{|ii\rangle}
\end{array}\end{equation}
be the maximally entangled state in the $d \times d$ system. It is
a stabilizer state and its stabilizer is $S=\langle
g_1,g_2\rangle$, where
\begin{equation}\begin{array}{l}
g_1=X_1X_2,\\
g_2=Z_1Z^{-1}_2.
\end{array}\end{equation}
The check matrix of $g_1,g_2$ is
\begin{equation}
M=\begin{pmatrix}
1 & 1 &  0 &  0\\
0 & 0 &  1 & -1\\
\end{pmatrix}
\end{equation}
Consider the partition $(\{1\},\{2\})$. The first and third
columns of $M$ are
$\overrightarrow{\alpha}_1=(1,0)^T,\overrightarrow{\beta}_1=(0,1)^T$.
Let $R_1=Q_1=\{1\}$. Then by theorem 1 and its proof, $d^2$ is an
achievable alphabet size for $|\psi_S\rangle$ with respect to the
bipartition $(\{1\},\{2\})$, and the corresponding encoding
operations are $\{\sigma_{i,j}:i,j=0,1,\dots,d-1\}$.
\end{example}

\begin{example}
Now consider the generalization of GHZ state to arbitrary
$n$-qudit system
\begin{equation}
|GHZ_{d,n}\rangle=\frac{1}{\sqrt{d}}\sum\limits_{i=0}^{d-1}{|ii\dots
i\rangle}.
\end{equation}
Its distributed dense coding capability has been investigated by
Refs. \cite{LL02,GW02,LA02}. One can see $|GHZ_{d,n}\rangle$ is a
stabilizer state and its stabilizer is $S=\langle
g_1,g_2,\dots,g_n \rangle$, where
\begin{equation}\begin{array}{ll}
g_1=X_1X_2\dots X_n, & \\
g_j=Z_{j-1}Z^{-1}_j, & \forall j=2,3,\dots,n.\\
\end{array}\end{equation}
The check matrix for $g_1,g_2,\dots,g_n$ is defined as follows:
$\forall i =1,2,\dots,n$, the $i$-th column is
\begin{equation}\begin{array}{l}
\overrightarrow{\alpha}_i=(1,0,0,\dots,0)^T;
\end{array}\end{equation}
the $(n+1)$-th column is
\begin{equation}\begin{array}{l}
\overrightarrow{\beta}_1=(0,1,0,\dots,0)^T;
\end{array}\end{equation}
$\forall i=2,3,\dots,n-1$, the $(n+i)$-th column is
\begin{equation}\begin{array}{l}
\overrightarrow{\beta}_i=(0,\dots,0,-1,1,0,\dots,0)^T,
\end{array}\end{equation}
where $-1$ and $1$ are the $i$-th and $(i+1)$-th entries of
$\overrightarrow{\beta}_i$ respectively; the $2n$-th column is
\begin{equation}\begin{array}{l}
\overrightarrow{\beta}_n=(0,0,\dots,0,-1)^T.
\end{array}\end{equation}

For example, when $n=4$, we have $g_1=X_1X_2X_3X_4$,
$g_2=Z_1Z^{-1}_2$, $g_3=Z_2Z^{-1}_3$, $g_4=Z_3Z^{-1}_4$. The
corresponding check matrix is
\begin{equation}
M=\begin{pmatrix}
1 & 1 & 1 & 1  & 0 &  0 &  0 &  0\\
0 & 0 & 0 & 0  & 1 & -1 &  0 &  0\\
0 & 0 & 0 & 0  & 0 &  1 & -1 &  0\\
0 & 0 & 0 & 0  & 0 &  0 &  1 & -1\\
\end{pmatrix}
\end{equation}

Now consider the partition $(\{1\},\{2\},\dots,\{n-1\},\{n\})$.
Define
\begin{equation}\begin{array}{l}
S_i=span\{\overrightarrow{\alpha}_i,\overrightarrow{\beta}_i\},
\end{array}\end{equation}
$\forall i =1,2,\dots,n-1$.

Choose $\overrightarrow{X}= (\overrightarrow{\alpha}_1,
\overrightarrow{\beta}_1, \overrightarrow{\beta}_2,\dots,
\overrightarrow{\beta}_{n-1})$ as the basis of $\mathbb{Z}^n_d$.
Then we obtain $P_1=\{1,2,\dots,n-1\}$, $P_j=\{j-1\}$, $\forall
j=2,3,\dots,n$. For $i=1,2,\dots,n-1$, $j \in P_i$, choose
$\overrightarrow{z}_{ij}$ as follows:
\begin{equation}\begin{array}{l}
\overrightarrow{z}_{i1}=\overrightarrow{\alpha}_i,\\
\overrightarrow{z}_{i(i+1)}=\overrightarrow{\beta}_i,\\
\end{array}\end{equation}
where $i=1,2,\dots,n-1$. Then $c_{i1}=c_{i(i+1)}=1$,
$i=1,2,\dots,n-1$.

Choose arbitrary
\begin{equation}
\overrightarrow{a}_1=(\lambda_1,\lambda_2,\dots,\lambda_{n-1}) \in
A(P_1;d),
\end{equation}
i.e. $\prod_{i=1}^{n-1}{\lambda_i} \le d$. Also, $\forall
j=2,3,\dots,n$, choose
\begin{equation}
\overrightarrow{a}_j=(1,\dots,1,d,1,\dots,1) \in A(P_j;d),
\end{equation}
where $d$ is the $(j-1)$-th entry of $\overrightarrow{a}_j$. Since
$\forall j=1,2,\dots,n$, $\forall i \in P_j$, $c_{ij}=1$, we have
$Q_j=P_j$. Consequently, $\forall j=1,2,\dots,n-1$,
\begin{equation}
\overrightarrow{b}_j=F(\overrightarrow{a}_{j};Q_j)=\overrightarrow{a}_j.
\end{equation}
By theorem 2, we have $(\lambda_1 d,\lambda_2
d,\dots,\lambda_{n-1} d)$ is an achievable alphabet size for
$|\psi_S\rangle$ with respect to
$(\{1\},\{2\},\dots,\{n-1\},\{n\})$. We realize that similar
results were also obtained by Refs. \cite{LL02,GW02}.
\end{example}

\begin{example}
Now consider a $5 \times 5 \times 5 \times 5 \times 5$ system,
i.e. $d=5$, $n=5$. Define
\begin{equation}\begin{array}{l}
g_1=\sigma_{1,1} \otimes \sigma_{1,2} \otimes \sigma_{2,1} \otimes \sigma_{0,1} \otimes \sigma_{3,0},\\
g_2=\sigma_{2,4} \otimes \sigma_{1,1} \otimes \sigma_{0,2} \otimes \sigma_{1,2} \otimes \sigma_{2,2},\\
g_3=\sigma_{2,2} \otimes \sigma_{4,1} \otimes \sigma_{1,1} \otimes \sigma_{0,1} \otimes \sigma_{3,2},\\
g_4=\sigma_{3,0} \otimes \sigma_{0,1} \otimes \sigma_{4,2} \otimes \sigma_{1,2} \otimes \sigma_{4,1},\\
g_5=\sigma_{4,3} \otimes \sigma_{1,4} \otimes \sigma_{2,3} \otimes \sigma_{4,2} \otimes \sigma_{2,3}.\\
\end{array}\end{equation}
They are five independent commuting operators in $G'^{(5)}_5$. Let
$S=\langle g_1,g_2,g_3,g_4,g_5\rangle$. The density matrix of the
state stabilized by $S$ is given by
\begin{equation}
\rho_S=\frac{1}{5^5}{\prod\limits_{i=1}^{5}{(\sum\limits_{j=0}^{4}{g^j_i}})}.
\end{equation}
The check matrix for $g_1,g_2,g_3,g_4,g_5$ is
\begin{equation}
M=\begin{pmatrix}
1 &   1  &  2  &  0 &   3  &   1  &   2 &  1  &  1 &   0\\
2 &   1  &  0  &  1 &   2  &   4  &   1 &  2  &  2 &   2\\
2 &   4  &  1  &  0 &   3  &   2  &   1 &  1  &  1 &   2\\
3 &   0  &  4  &  1 &   4  &   0  &   1 &  2  &  2 &   1\\
4 &   1  &  2  &  4 &   2  &   3  &   4 &  3  &  2 &   3
\end{pmatrix}.
\end{equation}

Consider the partition $(T_1,T_2,T_3)=(\{1,2\},\{3\},\{4,5\})$.
Suppose $A_1$ holds the first and second qudits, $A_2$ holds the
third qudit, and $A_3$ holds the fourth and fifth qudits. Let
$\overrightarrow{\alpha}_i$, $\overrightarrow{\beta}_i$ be the
$i$-th and $(i+5)$-th columns of $M$, $\forall i =1,2,3$. Define
$R_1=\{1,2\}$, $Q_1=\{2\}$, $R_2=\{3\}$, $Q_2=\{3\}$. One can
check that $\overrightarrow{\alpha}_1,\overrightarrow{\alpha}_2,
\overrightarrow{\alpha}_3,\overrightarrow{\beta}_2,\overrightarrow{\beta}_3$
are linearly independent. Therefore, by theorem 1 and its proof,
$\rho_S$ is optimally useful for deterministic distributed dense
coding with respect to $(\{1,2\},\{3\},\{4,5\})$. It can achieve
the alphabet size $(125,25)$ with the following protocol: $A_1$'s
encoding operations are $\{\sigma_{0,b_1} \otimes
\sigma_{a_2,b_2}:a_2,b_1,b_2=0,1,2,3,4\}$; $A_2$'s encoding
operations are $\{\sigma_{a_3,b_3}:a_3,b_3=0,1,2,3,4\}$.
\end{example}

\begin{example}
Now consider a $7 \times 7 \times 7 \times 7$ system, i.e. $d=7$,
$n=4$. Define
\begin{equation}\begin{array}{l}
g_1=\sigma_{1,3} \otimes \sigma_{2,2} \otimes \sigma_{2,0} \otimes  \sigma_{1,1},\\
g_2=\sigma_{1,5} \otimes \sigma_{1,1} \otimes \sigma_{3,2} \otimes  \sigma_{2,3},\\
g_3=\sigma_{2,3} \otimes \sigma_{1,0} \otimes \sigma_{4,5} \otimes  \sigma_{3,5},\\
g_4=\sigma_{3,1} \otimes \sigma_{0,0} \otimes \sigma_{6,6} \otimes
\sigma_{5,1}.
\end{array}\end{equation}
They are four independent commuting operators in $G'^{(4)}_7$. Let
$S=\langle g_1,g_2,g_3,g_4\rangle$. The density matrix of the
state stabilized by $S$ is given by
\begin{equation}
\rho_S=\frac{1}{7^4}{\prod\limits_{i=1}^{4}{(\sum\limits_{j=0}^{6}{g^j_i}})}.
\end{equation}
The check matrix for $g_1,g_2,g_3,g_4$ is
\begin{equation}
M=\begin{pmatrix}
1 &  2 &  2 &  1 &  3 &  2  &  0 &  1\\
1 &  1 &  3 &  2 &  5 &  1  &  2 &  3\\
2 &  1 &  4 &  3 &  3 &  0  &  5 &  5\\
3 &  0 &  6 &  5 &  1 &  0  &  6 &  1
\end{pmatrix}.
\end{equation}

Consider the partition $(\{1\},\{2\},\{3\},\{4\})$. Suppose
$A_1,A_2,A_3,A_4$ hold the first, second, third and fourth qudits
respectively.

Let $\overrightarrow{\alpha}_i,\overrightarrow{\beta}_i$ denote
the $i$-th and $(i+4)$-th columns of $M$, $\forall i=1,2,3$.
Define
\begin{equation}\begin{array}{l}
S_i=span\{\overrightarrow{\alpha}_i,\overrightarrow{\beta}_i\},
\end{array}\end{equation}
$\forall i=1,2,3$.

Choose the basis $\overrightarrow{X}=(\overrightarrow{x}_1,
\overrightarrow{x}_2,\overrightarrow{x}_3,\overrightarrow{x}_4)$
of $\mathbb{Z}^4_7$ as follows:
\begin{equation}\begin{array}{l}
\overrightarrow{x}_1=(0,0,0,1)^T,\\
\overrightarrow{x}_2=(0,0,1,0)^T,\\
\overrightarrow{x}_3=(0,1,0,0)^T,\\
\overrightarrow{x}_4=(1,0,0,0)^T.\\
\end{array}\end{equation}
Then we have $P_1=\{1,3\}$, $P_2=\{2,3\}$, $P_3=\{2\}$,
$P_4=\{1\}$.

For $j=1,2,3,4$, $i \in P_j$, choose $\overrightarrow{z}_{ij}$ as
follows:
\begin{equation}\begin{array}{l}
\overrightarrow{z}_{11}=\overrightarrow{\beta}_1=(3,5,3,1)^T,\\
\overrightarrow{z}_{14}=3\overrightarrow{\beta}_1-\overrightarrow{\alpha}_1=(1,0,0,0)^T,\\
\overrightarrow{z}_{22}=\overrightarrow{\alpha}_2=(2,1,1,0)^T,\\
\overrightarrow{z}_{23}=\overrightarrow{\beta}_2=(2,1,0,0)^T, \\
\overrightarrow{z}_{31}=-\overrightarrow{\beta}_3=(0,5,2,1)^T,\\
\overrightarrow{z}_{32}=\overrightarrow{\beta}_3-\overrightarrow{\alpha}_3=(5,6,1,0)^T.\\
\end{array}\end{equation}
Consequently, $c_{11}=c_{14}=c_{22}=c_{23}=c_{31}=c_{32}=1$.

Now choose
\begin{equation}\begin{array}{l}
(a_{11},a_{12},a_{13})=(2,1,3) \in A(P_1;7),\\
(a_{21},a_{22},a_{23})=(1,2,3) \in A(P_2;7),\\
(a_{31},a_{32},a_{33})=(1,7,1) \in A(P_3;7),\\
(a_{41},a_{42},a_{43})=(7,1,1) \in A(P_4;7).\\
\end{array}\end{equation}
Since $\forall j=1,2,3,4$, $\forall i \in P_j$, $c_{ij}=1$, we get
$Q_j=P_j$. Thus $b_{ji}=a_{ji}$, $\forall i=1,2,3$, $\forall
j=1,2,3,4$. By theorem 2, $(2\times 7,2 \times 7,3 \times
3)=(14,14,9)$ is an achievable alphabet size for $\rho_S$ with
respect to $(\{1\},\{2\},\{3\},\{4\})$. So $\rho_S$ is useful for
deterministic distributed dense coding with respect to this
partition. The corresponding dense coding protocol is built as
follows. $\forall j=1,2,3,4$, $\forall i \in P_j$, since
$c_{ij}=1$, we define
\begin{equation}\begin{array}{l}
\overrightarrow{y}_{ij}=\prod\limits_{t=1}^{i-1}{a_{jt}}\overrightarrow{z}_{ij}.
\end{array}\end{equation}
For all $i \not\in P_j$, define $\overrightarrow{y}_{ij}=0$. Then
by Eq.(\ref{equ:gammilambdai}),
\begin{equation}\begin{array}{l}
\overrightarrow{\gamma}(1;\overrightarrow{\lambda}_1)
={\lambda_{11}\overrightarrow{y}_{11}}+{\lambda_{14}\overrightarrow{y}_{14}}\\
=(\lambda_{11}+3\lambda_{14})\overrightarrow{\beta}_1-\lambda_{14}\overrightarrow{\alpha}_1,\\
\end{array}\end{equation}
where $\lambda_{11}=1,2$, $\lambda_{14}=1,2,\dots,7$;
\begin{equation}\begin{array}{l}
\overrightarrow{\gamma}(2;\overrightarrow{\lambda}_2)
={\lambda_{22}\overrightarrow{y}_{22}}+{\lambda_{23}\overrightarrow{y}_{23}}\\
=\lambda_{23}\overrightarrow{\beta}_2+\lambda_{22}\overrightarrow{\alpha}_2,\\
\end{array}\end{equation}
where $\lambda_{22} =1,2$, $\lambda_{23}=1,2,\dots,7$;
\begin{equation}\begin{array}{l}
\overrightarrow{\gamma}(3;\overrightarrow{\lambda}_3)
={\lambda_{31}\overrightarrow{y}_{31}}+{\lambda_{32}\overrightarrow{y}_{32}}\\
=2(-\lambda_{31}+\lambda_{32})\overrightarrow{\beta}_3-2\lambda_{32}\overrightarrow{\alpha}_3,\\
\end{array}\end{equation}
where $\lambda_{31},\lambda_{32} =1,2,3$. So $A_1$'s encoding
operations are
$\{\sigma_{\lambda_{11}+3\lambda_{14},\lambda_{14}}: \lambda_{11}
=1,2, \lambda_{14}=1,2,\dots,7\}$; $A_2$'s encoding operations are
$\{\sigma_{\lambda_{23},-\lambda_{22}}: \lambda_{22} =1,2,
\lambda_{23}=1,2,\dots,7\}$; $A_3$'s encoding operations are
$\{\sigma_{2(-\lambda_{31}+\lambda_{32}),2\lambda_{32}}:
\lambda_{31},\lambda_{32}=1,2,3\}$.
\end{example}

\section{Conclusion}

In sum, we have investigated the possibility of performing
deterministic distributed dense coding with the aid of a
previously shared stabilizer state. We present a sufficient
condition for a stabilizer state to be useful for deterministic
distributed dense coding with respect to a given partition plan.
The corresponding protocol is also constructed. Then a method is
proposed to partially solve the general problem of finding the set
of achievable alphabet sizes for an arbitrary stabilizer state
with respect to an arbitrary partition plan. Finally, our work
provides a new perspective from the stabilizer formalism to view
the standard dense coding protocol and also unifies several
previous results in a single framework.

We would like to point out several open questions that deserve
further research. The first question is whether one can achieve
the optimal alphabet sizes for any stabilizer state by using only
generalized Pauli group elements to encode. If so, can all the
optimal protocols be generated by our theorem 1 and theorem 2? The
second problem would be to consider deterministic distributed
dense coding with multiple copies of a stabilizer state. We do not
know whether the dense coding capacity of a stabilizer state can
be improved asymptotically. Finally, to our knowledge, there are
almost no results about \textit{deterministic} distributed dense
coding with a general multipartite entangled state. We hope our
results can shed light on the power of general multipartite
entanglement in this task.

\section*{Acknowledgement}
This work was partly supported by the Natural Science Foundation
of China (Grant Nos. 60621062 and 60503001) and the Hi-Tech
Research and Development Program of China (863 project) (Grant No.
2006AA01Z102).

\end{document}